\input pipi.sty
\input epsf.sty
\magnification1000
\raggedbottom

\nopagenumbers
\rightline{Revised 9 February 2004}
\rightline{FTUAM 03-19}
\rightline{hep-ph/0312187}

\bigskip
\hrule height .3mm
\vskip.6cm
\centerline{{\bigfib  Regge analysis of pion-pion (and pion-kaon) scattering}}
\medskip
\centerline{{\bigfib     for  energy $s^{1/2}>1.4$ GeV }}
\medskip
\centerrule{.7cm}
\vskip1cm
\setbox8=\vbox{\hsize65mm {\noindent\fib J. R. Pel\'aez} 
\vskip .1cm
\noindent{\addressfont Departamento de F\'{\i}sica Te\'orica,~II\hb
 (M\'etodos Matem\'aticos),\hb
Facultad de Ciencias F\'{\i}sicas,\hb
Universidad Complutense de Madrid,\hb
E-28040, Madrid, Spain.}}
\centerline{\box8}
\smallskip
\setbox7=\vbox{\hsize65mm \fib and} 
\centerline{\box7}
\smallskip
\setbox9=\vbox{\hsize65mm {\noindent\fib F. J. 
Yndur\'ain} 
\vskip .1cm
\noindent{\addressfont Departamento de F\'{\i}sica Te\'orica, C-XI\hb
 Universidad Aut\'onoma de Madrid,\hb
 Canto Blanco,\hb
E-28049, Madrid, Spain.}\hb}
\smallskip
\centerline{\box9}
\bigskip

\setbox0=\vbox{\abstracttype{Abstract} 
We perform a detailed Regge analysis of $NN$, $\pi N$, $KN$, $\pi\pi$ 
and $\pi K$ scattering.  
From it, we find expressions that represent 
 the $\pi\pi$ scattering amplitudes with an accuracy of a few percent, 
for exchange of isospin zero,
and 
$\sim15\%$ for exchange of isospin~1, and this
 for energies $s^{1/2}> 1.4\,{\rm GeV}$  and for momentum transfers
 $|t|^{1/2}\lsim 0.4\,{\rm GeV}$. 
These Regge formulas are perfectly compatible with the low energy
($s^{1/2}\sim 1.4\,{\rm GeV}$) scattering amplitudes 
deduced from  $\pi\pi$ phase shift analyses as well as 
 with higher energy ($s^{1/2}\gsim1.4\,{\rm GeV}$) experimental $\pi\pi$ cross sections. 
They are also  compatible with $NN$, $KN$ and $\pi N$ experimental cross sections 
using factorization, a property that we check with great precision. 
This contrasts with results from current phase shift analyses of 
the $\pi\pi$ scattering amplitude which  
bear little resemblance to reality in the region $1.4<s^{1/2}<2$~GeV, as they 
are not well defined and increasingly violate a number of physical requirements 
when the energy grows. 
$\pi K$ scattering is also considered, and we present a Regge analysis 
for these processes valid 
for energies  $s^{1/2}>1.7$ GeV.
\break\indent
As a byproduct of our analysis we present also a fit 
of $NN$, $\pi N$ and $KN$ cross sections 
valid from c.m. kinetic energy $E_{\rm kin}\simeq1$~GeV to multi TeV energies.
}
\centerline{\box0}
\brochureendcover{Typeset with \physmatex}
\brochureb{\smallsc j. r. pel\'aez and f. j.  yndur\'ain}{\smallsc 
regge analysis of  pion-pion   (and pion kaon) scattering }{1}

\brochuresection{1. Introduction}

\noindent
A precise and reliable knowledge of the $\pi\pi$ scattering amplitude 
has become increasingly important in the last years. 
This is so, in particular, because  $\pi\pi$ scattering is one of the few places where one has more
observables than unknown constants in  a chiral perturbation theory analysis, so it provides 
a window to higher order terms. 
Moreover, an accurate determination of the S wave scattering lengths, and of the phase shifts at
$s^{1/2}=m_K$, provide 
essential information for two subjects under intensive experimental investigation 
at present, viz., pionic atom decays
and  CP violation in the kaonic system.  In   recent papers, Ananthanarayan, Colangelo,
Gasser and Leutwyler\ref{1}  (that we will denote by ACGL),
 Colangelo, Gasser and Leutwyler,\ref{2}  Descotes et al.\ref{3} 
and  Kami\'nski,  Le\'sniak and Loiseau\ref{3} have used 
experimental information, analyticity and unitarity (in the form of  
the Roy equations) and, in ref.~2, chiral perturbation theory, 
to construct the $\pi\pi$ 
scattering amplitude at low energy, 
$s^{1/2}\leq0.8\,\gev$. 
For these analyses one needs as input the imaginary part of the $\pi\pi$ amplitudes above the
energy 
at which the Roy analysis stops; 
in particular, one needs the scattering amplitudes for $s^{1/2}$ above 1.4 \gev, 
which will be the subject of the present paper.

Unfortunately, the authors in refs.~2,~3 take their 
 $\pi\pi$ scattering
amplitude in this energy region from  ACGL,\ref{1} which 
presents a number of serious  drawbacks.\fnote{
In ref.~4, the Regge parameters of ACGL are also used for
$\pi K$  scattering; perhaps this is the reason why they are not able to get 
a satisfactory description of this process.}   First of all, the input scattering
amplitude at  energy $s^{1/2}\gsim 2\,\gev$ which these authors use 
 (following Pennington\ref{5}) is not physically acceptable, as 
it contradicts known properties of standard Regge theory
 and, moreover, is 
quite incompatible with {\sl experimental\/}\fnote{It should be noted
 that Pennington has publicly stated 
(in the Conversano workshop, July 2003) that 
his analysis, tenable in 1974, is superseded by more recent developments, 
both experimental and theoretical. 
In fact, already by 1977 it was clear to experts that 
standard Regge behaviour also holds for $\pi\pi$ scattering; see, e.g., 
Froggatt and Petersen, ref.~6,
who use the correct Regge behaviour in their dispersive analysis of 
$\pi\pi$ scattering.}
 $\pi\pi$ total cross sections,\ref{7} and this in spite of the 
large errors assumed by ACGL. 
Secondly, the scattering amplitude for $1.4\,\gev\leq s^{1/2}\leq1.9\,\gev$ that 
ACGL (and, following them, the authors in refs.~2,~3) use 
 is obtained from phase shift analyses, 
specifically the Cern--Munich set of analyses,\ref{8} 
which are subject to large uncertainties and which, indeed,  can be shown to
contradict a number of physical requirements. [Although we 
will not discuss this here (see ref.~9), it is also clear that the errors ACGL, and the authors in
ref.~2,  take  for some of their lower energy
 experimental input data are  excessively optimistic and, 
moreover,  
 certain of their chiral parameters are likely to be biased\ref{10}]. 
One should imagine that the use of incorrect high energy input should lead 
to inconsistent low energy output. In fact, this occurs in the work 
by  Colangelo, Gasser and Leutwyler,\ref{2} where the 
central values are probably displaced and the errors claimed are 
excessively optimistic and lead to several mismatches, as shown in refs.~9,~11.

In the present note we will not concern ourselves with the 
reliability or otherwise of the {\sl low energy} consequences 
of faulty high energy input, but will concentrate our efforts in 
ascertaining what a {\sl correct}  
high energy input should be. 
To do this, we  will 
 perform a detailed Regge analysis and show that it   
is  compatible with {\sl experimental} data for all 
values of $s^{1/2}\gsim1.4\,\gev$ (for some $\pi\pi$ 
processes, even down to $s^{1/2}\sim1\,\gev$). 
The resulting $\pi\pi$ amplitudes,  summarized in Eqs.~(4), (5), (11) and (18) and
Table~II   
below, should provide a correct and accurate input for  dispersive studies of 
$\pi\pi$ scattering. 

Our analysis will be an improvement on standard ones not only for 
$\pi\pi$ and $\pi K$, but even for 
$\pi N$, $KN$ and $NN$ in that we will be able to
 give an accurate description of the amplitudes for energies ranging from 
a kinetic enrgy in the center of mass 
$E_{\rm kin}\simeq1\,\gev$ to the TeV region. 
This accuracy reaches the level of a very few percent 
for zero isospin exchange, and it is less precise 
for the isospin~1 exchange amplitude, for which the errors may go up to $\sim15\%$ at low energy. 

An analysis of high energy $\pi K$ scattering is possible by a straightforward extension of the 
methods  here; 
it is given in \sect~3, where we present   precise 
Regge formulas for zero isospin exchange, valid for energies $s^{1/2}>1.7\,\gev$. 

The analysis of $\pi\pi$, $\pi K$ scattering up to (relatively) low energies, $\sim 14\,\gev$, 
is described in \sects~2,~3; in \sect~4 we extend it to 
multi~TeV energies. 
As a byproduct of our
analysis we present also a parametrization   of  $NN$, $\pi N$ and $KN$ total
cross sections  compatible with the Froissart bound and 
valid from  $E_{\rm kin}\simeq1$~GeV to $\sim30$~TeV.
In particular, we predict the total $pp$ cross section at the LHC to be 
$$\sigma_{pp}=116\pm4\;{\rm mb}.$$

Our results are summarized in \sect~5, where a brief discussion is also presented.

\booksection{2.  Regge analysis of $\pi\pi$ scattering ($s^{1/2}\geq 1.4\,\gev$)}

\noindent
 We normalize scattering amplitudes to
$$\sigma_{AB}=
\dfrac{4\pi^2}{\lambda^{1/2}(s,m^2_A,m^2_B)}\,\imag F_{A+B\to A+B}(s,0);
\quad \lambda(a,b,c)=a^2+b^2+c^2-2ab-2ac-2bc.
$$
$\sigma_{AB}$ is the total $A+B$ cross section; 
for $NN$ ($\bar{p}p$, $pp$) and $\pi N$ scattering, we understand that the cross sections are
spin-averaged. According to Regge theory, the 
imaginary part of a scattering amplitude 
with fixed isospin in the $t$ channel, $\imag F^{(I_t)}_{A+B\to A+B}(s,t)$, 
{\sl factorizes}\fnote{In potential theory 
factorization can be proved rigorously; in relativistic theory, it follows 
from extended unitarity or, in QCD, from the DGLAP formalism.\ref{12}} as a product: 
for each Regge pole, $R$, we can write
$$\imag F^{(I_t)}_{A+B\to A+B}(s,t)\simeqsub_{{s\to\infty}\atop{t\,{\rm fixed}}}
f_A^{(R)}(t)f_B^{(R)}(t)(s/\hat{s})^{\alpha_R(t)}.
\equn{(1a)}$$ 
Here $\hat{s}$ is a constant, usually taken to be $1\,\gev^2$; we will do so here. 
A similar formula holds for the real parts:
$$\real F^{(I_t)}_{A+B\to A+B}(s,t)\simeqsub_{{s\to\infty}\atop{t\,{\rm fixed}}}
\real\xi(R)\,f_A^{(R)}(t)f_B^{(R)}(t)(s/\hat{s})^{\alpha_R(t)}.
\equn{(1b)}$$
 $\xi(R)$, with  $\imag\xi(R)=1$, is known as the  {\sl signature} factor; for the Pomeron ($P$),
$P'$ and rho Regge poles one has
$$\real\xi(R)=-\dfrac{1+\cos\pi\alpha_R}{\sin\pi\alpha_R},\;R=P,\,P';
\quad \real\xi(\rho)=\dfrac{1-\cos\pi\alpha_\rho}{\sin\pi\alpha_\rho}.
\equn{(1c)}$$
 The residue functions $f^{(R)}_i(t)$ depend on the 
quantum numbers of the Regge pole exchanged, on the  particles 
that couple to it and, if we had  
external currents, also on their virtuality;   
but the power $(s/\hat{s})^{\alpha_R(t)}$ is universal and
 depends only on the Regge pole exchanged in 
channel $t$. 
The exponent $\alpha_R(t)$ is  the Regge trajectory associated to the 
quantum numbers in channel $t$. For the Pomeron, which is rather flat,
we will take it linear; for the rho, a more precise quadratic formula may be used. 
We thus write, for small $t$, 
$$\eqalign{
\alpha_P(t)\simeqsub_{t\sim0}\alpha_P(0)+\alpha'_Pt,\quad
\alpha_\rho(t)\simeqsub_{t\sim0}\alpha_\rho(0)+\alpha'_{\rho}\,t+
\tfrac{1}{2}\alpha''_{\rho}\,t^2.\cr
}
\equn{(2)}$$
For the $\rho$ and Pomeron pole, fits to high energy $\pi N$ and $NN$ processes give
$$\eqalign{
\alpha_\rho(0)=&\,0.52\pm0.02,\quad\alpha'_\rho=
0.90\, {\gev}^{-2},\quad \alpha''_{\rho}=-0.3\;{\gev}^{-4};\cr
\alpha_P(0)=&\,1,\quad\alpha'_P=0.2\pm0.1\, {\gev}^{-2}.\cr
}
\equn{(3)}$$
The 
Regge parameters taken here are based those in the global
 fit of  Rarita et al.,\ref{13} which are still the best
available  as there  are few modern data for the {\sl slopes}
 in the relevant energy range. 
There are a few differences, however. 
For $\alpha_\rho(0)$, 
 we take the value $0.52\pm0.02$, instead of 0.58. 
  This is more consistent with determinations
 based on deep inelastic scattering (see e.g.  the paper of Adel et al., in ref.~12).
Moreover, for 
$\alpha_\rho(t)$ 
we use a quadratic formula that agrees with the average slope of ref.~13 for small,
negative $t$, and which 
fulfills the condition $\alpha_\rho(M^2_\rho)=1$. 
Finally, for $\alpha'_P$, Rarita et al. give $0.11$, 
Froggatt and Petersen\ref{6} $0.3$ and the shrinking of the 
diffraction peak at the Tevatron suggests $0.26$. 
Our choice here encompasses these three values.  
These are minor improvements as, in fact, for our fits 
in the present paper we only need 
the values of the $\alpha_R(0)$; the slopes only intervene in sum rules.

Let us now turn to the functions $f_i(t)$. With respect to them we have two 
quite separate questions. 
First of all, we have the question of their normalization, that is 
to say, the values $f_i(0)$. 
These can be obtained with little ambiguity and small errors 
 by fitting experimental $NN$, $\pi N$ and $\pi\pi$ 
total cross section data; 
we will do precisely that below. 
A different matter is the dependence of the $f_i(t)$ on $t$, i.e., 
the ratios $f_i(t)/f_i(0)$, which is important in particular for Roy equations 
or sum rules like the ones at the end of the present Section. 
These are obtained from fits to the slopes of $NN$, $\pi N$ differential 
cross sections. Unfortunately,   
these fits are not unique, because both the background 
and the functional forms assumed for the $f_i(t)$ 
have a nonnegligible influence on the results, and because for the 
differential cross sections also the {\sl real} part of 
the scattering amplitudes intervene. 
Moreover, the parameters of these fits were obtained before QCD emerged as the theory of 
strong interactions; these fits  were extended to large  
values of $t$ where, as we now know, Regge theory must fail  and one has instead 
the Brodsky--Farrar behaviour.\ref{14} They are thus forced fits. 

The situation, however, is not hopeless; the difference between the {\sl numerical} 
results of various fits is small, for  small values of $|t|$. For example, the
 numerical difference for the ratios
  $f_P(t)/f_P(0)$ between refs.~9,~16 
is below the $10\%$ level for $|t|^{1/2}\leq0.4\,\gev$, which covers the 
values of $t$ in which we are interested here. 
In the present paper we have chosen the $t$ dependence of ref.~13, 
which was obtained in a detailed fit to many data.

Before writing explicit formulas for the various processes ($NN$, $\pi N$, $\pi\pi$) 
we have to decide in which variable we assume Regge behaviour to hold, 
which is important for us since we are going down to rather low energies. 
In (1) we have taken the c.m. energy squared, 
$s=(p_1+p_2)^2$, with $p_i$ the momenta of the incoming particles. 
Other possibilities are the s-u crossing symmetric variable,
 $\nu=2p_1\cdot p_2$, and $E^2_{\rm kin}$, 
 so we could assume behaviours like $\nu^{\alpha_P}$ or
$E^{2\alpha_P}_{\rm kin}$
 instead of $s^{\alpha_P}$, etc. 
We have, in our fits, tried all three possibilities; the fits using $s$, as in (1), all have 
substantially better \chidof\ than those using  $\nu=2p_1\cdot p_2$ or $E^2_{\rm kin}$. 
Therefore, we stick to Regge behaviour in the variable $s$, as in (1).

\medskip
\noindent{\sl Regge formulas for $\pi\pi$, $\pi N$ and $NN$ scattering.}\quad
We  start with 
$\pi\pi$ scattering. For  exchange of isospin $I_t=0$ in the $t$ 
channel, containing the Pomeron and $P'$ pole (the second associated 
with the $f_2(1270)$ resonance) 
we have
$$\eqalign{ 
\imag F^{(I_t=0)}_{\pi\pi}(s,t)&\,\simeqsub_{{s\to\infty}\atop{t\,{\rm fixed}}}
P(s,t)+P'(s,t),\cr
P(s,t)=&\,\beta_P\,\alpha_P(t)\,\dfrac{1+\alpha_P(t)}{2}\,
\ee^{bt}(s/\hat{s})^{\alpha_P(t)},\cr
P'(s,t)=&\,\beta_{P'}\,\dfrac{\alpha_{P'}(t)[1+\alpha_{P'}(t)]}{\alpha_{P'}(0)[1+\alpha_{P'}(0)]}\,
\ee^{bt}(s/\hat{s})^{\alpha_{P'}(t)},\quad
\alpha_{P'}(t)=\alpha_\rho(t);\cr
b=&\,(2.4\pm0.2)\,{\gev}^{-2}.\cr
}
\equn{(4a)}$$
Here $\beta_P=[f^{(P)}_\pi]^2$, $\beta_{P'}=[f^{(P')}_\pi]^2$.
 
The expression (4a) is like its counterpart in ref.~13, except 
for the $P'$ pole parameters. In fact, 
the subleading contribution
of the  $P'$ pole, that is 
necessary at the lowest energy range, is added somewhat empirically; 
its parameters are not well known, and we simply assume the corresponding trajectory to be
degenerate 
with the one of the rho, as is suggested by a number of theoretical developments (in particular 
the QCD theory of Regge trajectories\ref{12}), 
and as is done in ref.~6: $\alpha_{P'}(t)=\alpha_\rho(t)$.   
 In ref.~13, a larger value (0.7
instead of 0.52) was given for the intercept of the $P'$ pole
  and a smaller number  was taken for
its residue, but more modern determinations, as well as our fits, substantiate
 our choice; see \sect~4, where we will 
present  a global fit to data leaving, in particular, 
$\alpha_{P'}(0)$ as a free parameter. 
The result for it, $\alpha_{P'}(0)=0.55\pm0.03$, 
is in perfect agreement with other modern determinations and with 
the degeneracy assumption.

It should perhaps also be remarked that Eq.~(4a), in what respects 
the Pomeron, is of limited validity (up to 10 -- 15 \gev) 
since, at higher energies, total cross sections are known to rise. 
A modification of $P(s,t)$ in (4a) that 
will make the parametrization valid up to multi-TeV energies 
will be given in \sect~4.

 For $I_t=1$, we  also take the parametrization of  ref.~13. 
We write
$$\eqalign{
\imag F^{(I_t=1)}_{\pi\pi}(s,t)&\,\simeqsub_{{s\to\infty}\atop{t\,{\rm fixed}}}
\rho(s,t);\cr
\rho(s,t)=&\,\beta_\rho\,\Big[(1.5+1)\ee^{bt}-1.5\Big]
\dfrac{1+\alpha_\rho(t)}{1+\alpha_\rho(0)}\,
(s/\hat{s})^{\alpha_\rho(t)}.\cr
}
\equn{(4b)}$$
$b$ is as before and  $\beta_\rho=[f^{(\rho)}_\pi]^2$.  The universal value of the slope of the 
diffractive factor, $\ee^{bt}$, for all three trajectories rho, $P$ and $P'$, is what 
was found in ref.~13 from fit to actual $NN$ and $\pi N$ data; 
it can nowadays be understood physically as a consequence of the universality 
of the Regge mechanism in QCD. 
We note that Froggatt and Petersen,\ref{6} who fit $\pi^+\pi^-$ data, find a value for $b$ 
similar to ours for the Pomeron, 
but somewhat different ones for rho and $P'$. 
This last fact is not very meaningful as, in the fits to $\pi^+\pi^-$, 
the $\rho$, $P'$ Regge poles are subleading and easily hidden by the Pomeron. 
We also remark that, in ref.~11, we had added a small background to $\imag F^{(I_t=1)}_{\pi\pi}$
to join smoothly the asymptotic formulas to the experimental cross section at 
$s^{1/2}\sim1.4\,\gev$.
 With the value of the parameter $\beta_\rho$ found in the 
present Section, such background is unnecessary.

For $\pi\pi$ scattering 
we have to add an amplitude for exchange of isospin 2, corresponding to double rho exchange,
which we do by writing
$$\imag F^{(I_t=2)}_{\pi\pi}(s,t)\simeqsub_{{s\to\infty}\atop{t\,{\rm
fixed}}}R_2(s,t)\equiv
\beta_2\ee^{bt}(s/\hat{s})^{\alpha_\rho(t)+\alpha_\rho(0)-1}.
\equn{(5)}$$ 
We will discuss this quantity  $R_2(s,t)$ later on; in 
particular, we will determine the quantity  $\beta_2$, which is small. 
We will start by putting  $\beta_2=0$ 
and correct for this afterwards.

The important parameters are $\beta_P$, $\beta_{P'}$, $\beta_\rho$. 
We can obtain them fitting $NN$ ($pp$ and $\bar{p}p$) and $\pi N$  
cross sections (including the forward differential cross section for the 
charge exchange reaction $\pi^- p\to\pi^0 n$), 
from $\pi\pi$ cross sections or from a global fit to the two sets. 
We write
$$\eqalign{
\dfrac{\sigma_{pp}+\sigma_{\bar{p}p}}{2}
\simeqsub_{s\;{\rm large}}&\;\dfrac{4\pi^2}{\lambda^{1/2}(s,m_p^2,m_p^2)}
\,\tfrac{1}{2}f_{N/\pi}^2\Big[P(s,0)+P'(s,0)\Big],\cr
\sigma_{\pi^\pm p}
\simeqsub_{s\;{\rm large}}&\;\dfrac{4\pi^2}{\lambda^{1/2}(s,m^2_\pi,m_p^2)}f_{N/\pi}
\left\{\dfrac{1}{\sqrt{6}}\Big[P(s,0)+P'(s,0)\Big]\mp
\tfrac{1}{2}\bar{\rho}(s,0)\right\},\cr
\left.\dfrac{\dd\sigma(\pi^- p\to\pi^0 n)}{\dd t}\right|_{t=0}
\simeqsub_{s\;{\rm large}}&\;f_{N/\pi}^2\,
\dfrac{1-\cos\pi\alpha_\rho}{\sin^2\pi\alpha_\rho}\,
\dfrac{\pi^3}{\lambda(s,m^2_\pi,m_p^2)}\,\left|\bar{\rho}(s,0)\right|^2.
\cr}
\equn{(6a)}$$
Here $f_{N/\pi}\equiv f_N^{(P)}/f_\pi^{(P)}$, and  
we have defined
$$\bar{\rho}(s,t)=\beta^{(N\pi)}_\rho\,\Big[(1.5+1)\ee^{bt}-1.5\Big]
\dfrac{1+\alpha_\rho(t)}{1+\alpha_\rho(0)}\,
(s/\hat{s})^{\alpha_\rho(t)}
\equn{(6b)}$$
with
$$\beta^{(N\pi)}_\rho=\left[f^{(P)}_\pi f^{(\rho)}_N/ f^{(\rho)}_\pi f^{(P)}_N\right]\beta_\rho.
\equn{(6c)}$$
In Eq.~(6a) we have put the same values of $f_{N/\pi}$ for Pomeron and 
$P'$. 
In \sect~4 we will discuss fits allowing for different
 $f^{(P)}_{N/\pi}$,  $f^{(P')}_{N/\pi}$; their central values 
will be somewhat different, but the improvement in the \chidof\ 
obtained by so doing is not significative.

\topinsert{
\medskip
\setbox3=\vbox{\hsize 10.5truecm\epsfxsize 9truecm\epsfbox{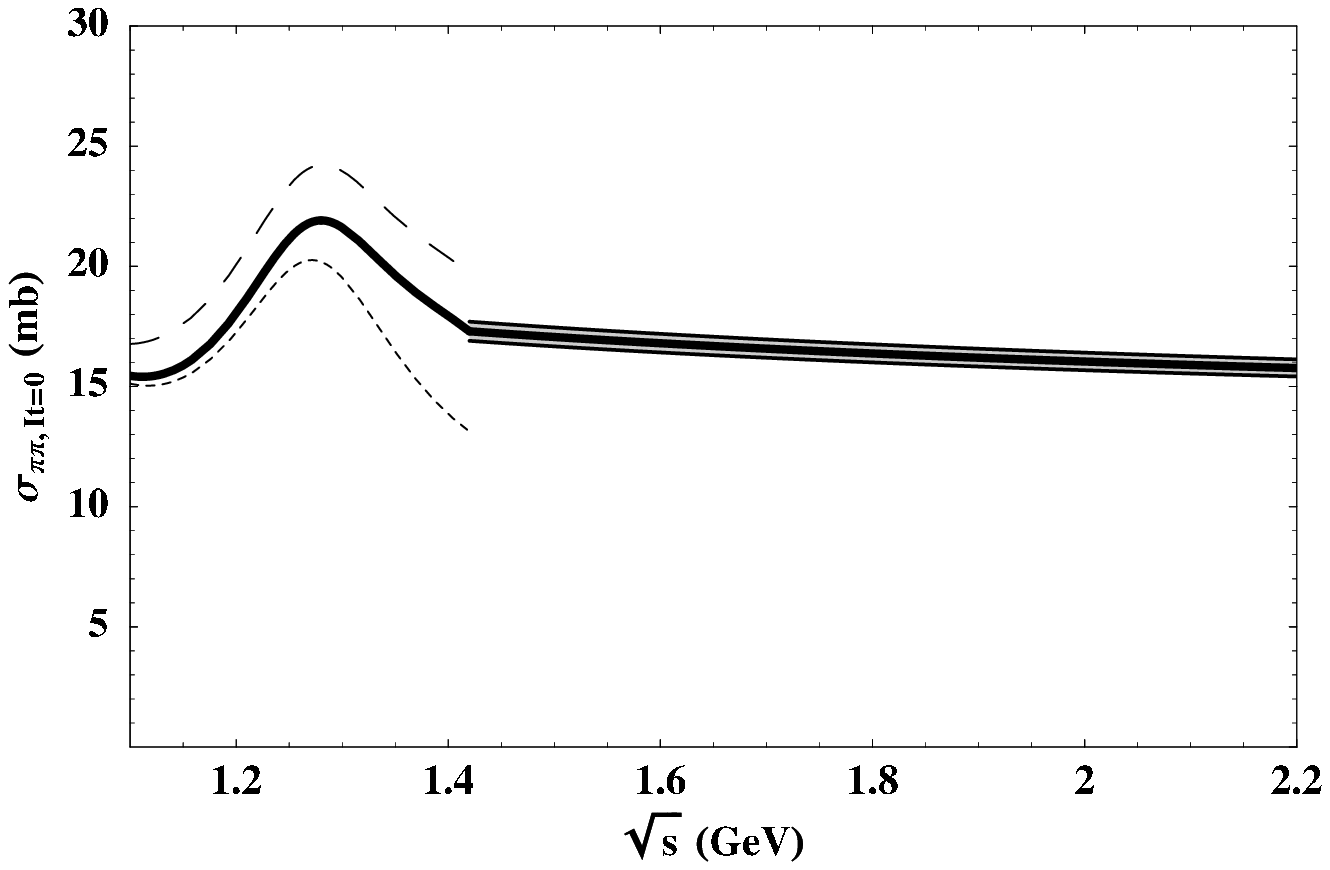}\hfil}
\setbox5=\vbox{\hsize 5.5truecm\captiontype\figurasc{Figure 1. }
 {The average cross section $\tfrac{1}{3}[2\sigma_{\pi^0\pi^+}+\sigma_{\pi^0\pi^0}]$, 
which is pure $I_t=0$.  
 Continuous lines, for $s^{1/2}>1.4\,\gev$: Regge formula. The 
 lines   cover the errors in the values of the 
Regge residues. Continuous lines, up to $s^{1/2}=1.4\,\gev$:  
experimental cross section (from the fits in 
ref.~11; actually, with a slightly improved 
D2 wave).
 The dotted and dashed lines are representative of the 
experimental errors in the  cross section.}}
\line{{\box3}\hfil\box5}
\bigskip
}\endinsert

\smallskip
\noindent{\sl Fits.}\quad  
We will not fit data for scattering off neutrons which would not improve the precision while, 
because the neutrons are necessarily bound, they could distort the fits. 
We will also not include the difference of cross sections $\sigma_{\bar{p}p}-\sigma_{pp}$ 
in the fits, as this would involve the contribution of at least three 
Regge poles ($\omega$, $\phi$ and $\pi$) which do not contribute to 
$\pi\pi$. One could include the reaction $\bar{p}p\to\bar{n}n$, which only involves 
exchange of the rho, but the data for it are few and with 
(comparatively) large errors, so it would add little to 
the analysis.
For the charge-exchange reaction, $\pi^- p\to\pi^0 n$, only data
 in the forward direction are included. 
This reaction is interesting in that, although it has much larger 
errors than the others, it receives contribution from the real part of the 
corresponding Regge pole, so it represents a completely independent test 
of the Regge formulas.

Before going on to the actual fits, a few words have to be said
  on the energy regions in which one may
expect Regge behaviour (and, in particular, factorization) to hold. 
Generally speaking, we expect this to occur when one is past the region of elastic resonances 
and one also has $E^2_{\rm kin}\gg\lambdav^2$ ($\lambdav\simeq0.4\,\gev$ is 
the QCD parameter),  
which means for $E_{\rm kin}\gsim1\,\gev$; but the 
precise details vary for different reactions. 
Thus, for $pp$, $\bar{p}p$ scattering 
there are no resonances and hence Regge behaviour is expected to occur 
precociously: 
here we will actually fit from $E_{\rm kin}=0.98\,\gev$.

For $\pi\pi$ scattering it is difficult to tell when exactly one may use Regge formulas 
since data, particularly for $\pi^-\pi^-$, are not very good. 
For the cross section
 $\sigma^{(I_t=0)}\equiv\tfrac{1}{3}[2\sigma_{\pi^0\pi^+}+\sigma_{\pi^0\pi^0}]$,
Eq.~(4) provides a good representation  for energies as low as $E_{\rm kin}=1\,\gev$, 
as shown in \fig~1; 
but, when resonances are more important, Regge behaviour
 is a good approximation only at slightly  
  higher energies.  
Another matter is that, at low energies ($s^{1/2}\sim1.5\,\gev$) the
$\pi\pi$ data are of poor quality. 
Because of this, 
we will consider two extreme possibilities for actual fits. 
The first, that we will call {\sl no-cut}, consists in including all 
$\pi\pi$ data for $E_{\rm kin}>1.1\,\gev$ ($s^{1/2}\geq1.38\,\gev$). 
The second possibility, that we call {\sl cut}, consists in cutting out all data for 
energies below $s^{1/2}=2$ \gev. 
The difference in results between the two fits will be an indication of the 
{\sl systematic} errors in our calculation.

\midinsert{
\setbox3=\vbox{\hsize16truecm{\epsfxsize 14.0truecm\epsfbox{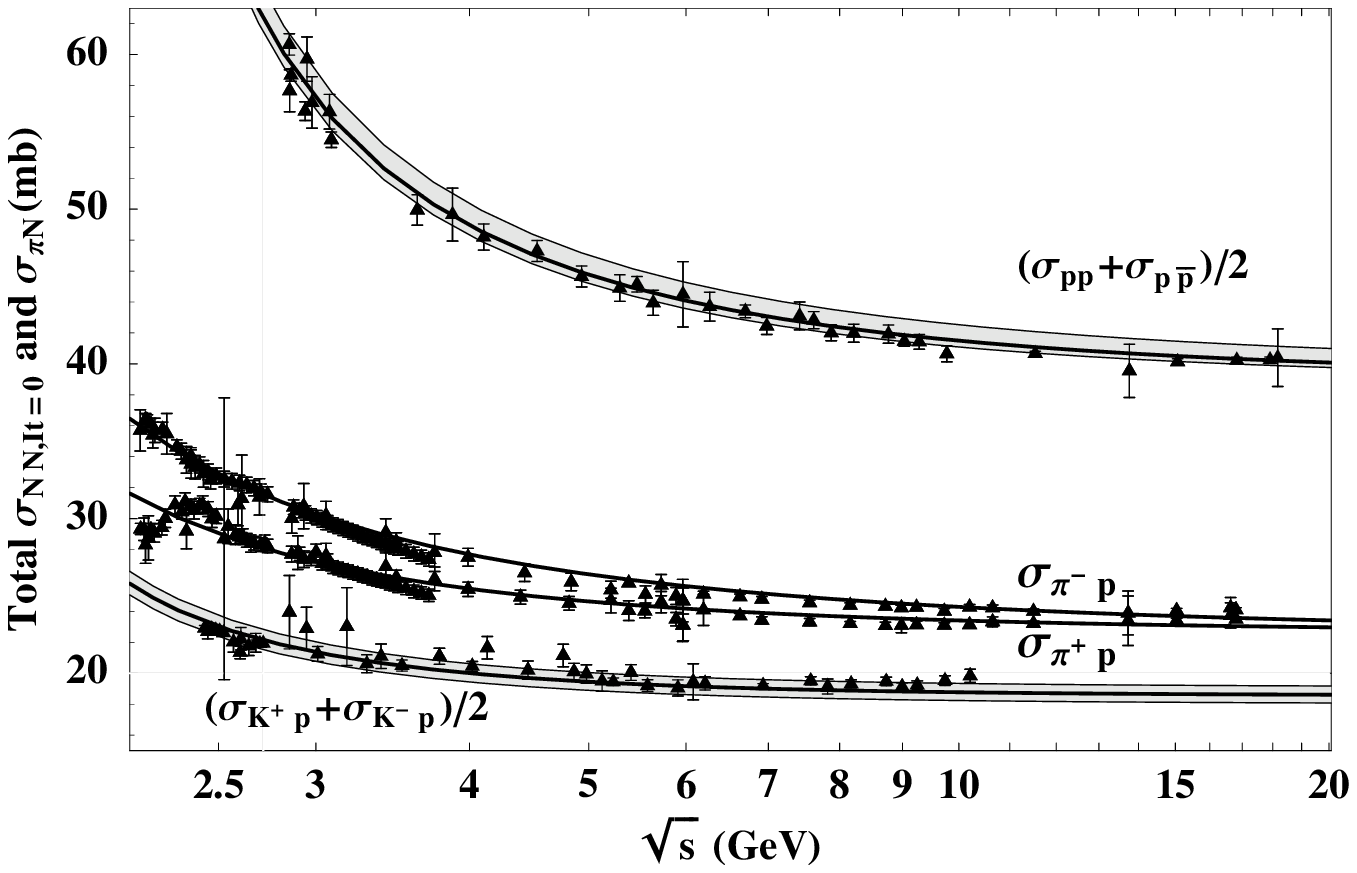}}}
\centerline{{\box3}\hfil}
\bigskip
\setbox5=\vbox{\hsize 14.5truecm\captiontype\figurasc{Figure 2. }
 {The total cross sections
$\sigma_{\pi^\pm p}$, $\tfrac{1}{2}(\sigma_{\bar{p}p}+\sigma_{pp})$ and 
$\tfrac{1}{2}(\sigma_{K^+p}+\sigma_{K^-p})$. 
Black dots, triangles and squares: experimental points.
Continuous lines: Regge formulas, with parameters as in 
our best fit.
For $\tfrac{1}{2}(\sigma_{\bar{p}p}+\sigma_{pp})$
 and $\tfrac{1}{2}(\sigma_{K^+p}+\sigma_{K^-p})$,
the three lines  cover the errors in the  values of the 
Regge residues. For $\pi N$ the theoretical error is of the order of that for 
 $\tfrac{1}{2}(\sigma_{\bar{p}p}+\sigma_{pp})$. Note that the thick line 
in the low energy experimental cross sections for $\pi N$ is merely due to the 
accumulation of closely spaced data.}}
\centerline{\box5}
}\endinsert

For $\pi N$ the formulas (6) fit well data down to 
$E_{\rm kin}\sim1.3\,\gev$, but, for the sum, $\sigma_{\pi^+ p}+\sigma_{\pi^- p}$, 
one can go to $E_{\rm kin}\sim1\,\gev$. For the difference,
 $\sigma_{\pi^+ p}-\sigma_{\pi^- p}$ and for the charge-exchange reaction 
$\pi^- p\to\pi^0 n$, resonances somewhat spoil {\sl local} 
agreement, but Eq.~(6) provides a good {\sl average} representation 
even down to 1 \gev, as has been known for a long time 
(see, e.g., ref.~15) and as can be seen in the lower 
energy region in our fit to $\pi^+p$ data in 
\fig~2. We will here 
start from $E_{\rm kin}=1.08\,\gev$.

Another question is how high one goes in energy.
In the present Section we fit  experimental data for c.m. kinetic energies  
$ E_{\rm kin}\lsim16.5$ \gev: 
this is what is required for applications to $\pi\pi$ Roy equations, 
dispersion relations and sum rules, 
since here the importance of 
 the very high energy region is negligible. 
Nevertheless, and as stated before, 
parametrizations and fits valid up to multi TeV energies will be given in  
\sect~4. 

 The data on $\pi^-p\to\pi^0n$ are from the 
compilation in ref.~15. 
For $NN$ and $\pi N$ we will take the data from the COMPAS Group
compilations,  as given in 
 the Particle Data Tables.\ref{16} 
For those 
data where systematic errors are not given, we have included a common systematic error of 
0.5\% for $pp$, $1\%$ for $\bar {p}p$ and $1.5\%$ for $\pi p$,
 which are like the standard systematic errors in other
data. 
Another possibility is to take a common systematic error of 1.5\% for all data: 
the difference of the results with the two will indicate the systematic errors of our fit.
We have also made  
 a selection  of $NN$, $\pi N$ data, as follows. 
 We  take only data at energies at which there
are  results for both $pp$ and $\bar{p}p$ or $\pi^+ p$ and $\pi^- p$; and, when 
there are, at a given energy, data from various experiments, we have taken only the most recent.
 This is designed to thin out the data to a number 
comparable in order of magnitude to that of  
$\pi\pi$, so that $\pi\pi$ data have a 
nonnegligible weight in the joint fits.
 For $\pi\pi$ scattering we 
have taken the errors as given by the various 
experimental groups except for those of 
Abramowicz et al.,\ref{7} who only 
give statistical errors, much smaller than those of the other groups, and  
for which we have added  a common systematic error of 1.5~mb to all points; 
even with this, the error, though comparable, is smaller than what other groups find.

We could fit separately the $NN$, $\pi N$ data and the $\pi\pi$ data of ref.~7,
 or make a global fit.  The results of these fits, in which we have put  $\beta_2=0$, 
and fixed $\alpha_\rho(0)=0.52$, 
 are given in Table~I, where the errors correspond to one standard deviation. 
The best values are average values, with errors enlarged to overlap  other results.
A graphical representation of this best fit may be seen, compared with experimental 
$NN$, $\pi N$ cross sections in \fig~2, and for  $\pi\pi$ data, 
in \fig~3. We 
note that, in \fig~3 for $\pi\pi$, we have used the values 
of $\beta_\rho$ and $\beta_2$ from Eq.~(11) below.

\midinsert{
\vskip0.5truecm
\setbox0=\vbox{
\setbox1=\vbox{\petit \offinterlineskip\hrule
\halign{
&\vrule#&\strut\hfil\ #\ \hfil&\vrule#&\strut\hfil\ #\ \hfil&
\vrule#&\strut\hfil\ #\ \hfil&
\vrule#&\strut\hfil\ #\ \hfil&\vrule#&\strut\hfil\ #\ \hfil\cr
 height2mm&\omit&&\omit&&\omit&&\omit&&\omit&\cr 
&\hfil \hfil&&\hfil $NN,\,\pi N$ [enlarged error$^{\rm (a)}$] \hfil&
&Only $\pi\pi$  [cut$^{\rm (b)}$  $^{\rm (c)}$]&
&\hfil $NN,\,\pi N,\,\pi\pi$ [cut$^{\rm (b)}$]\hfil&
&\hfil Best values\hfil& \cr
 height1mm&\omit&&\omit&&\omit&&\omit&&\omit&\cr
\noalign{\hrule}
height1mm&\omit&&\omit&&\omit&&\omit&&\omit&\cr
&$f_{N/\pi}$&&\vphantom{\Big|}$1.405\pm0.005\;[1.411\pm0.004] $&&$ $&
&\hfil$1.404\pm0.005\;[1.407\pm0.005]$ \hfil&&
$1.406\pm0.007$& \cr 
\noalign{\hrule}
height1mm&\omit&&\omit&&\omit&&\omit&&\omit&\cr
&\vphantom{\Big|}$\beta^{(N\pi)}_\rho$ 
\phantom{\big|}&&\phantom{\Big|}$0.366\pm0.009\;[0.367\pm0.010]$&&$ $&
&\hfil $0.366\pm0.010\;[0.367\pm0.009]$   \hfil&&\hfil 
$0.366\pm0.010$& \cr 
height1mm&\omit&&\omit&&\omit&&\omit&&\omit&\cr
\noalign{\hrule}  
height1mm&\omit&&\omit&&\omit&&\omit&&\omit&\cr
&\vphantom{\Big|}$\beta_\rho$ 
\phantom{\big|}&&\phantom{\Big|}$ $&&$1.30\pm0.13\;$[$0.59\pm0.27$]&
&\hfil $1.32\pm0.13\;[0.59\pm0.25]$   \hfil&&\hfil 
$1.0\pm0.3^{(*)}$& \cr 
height1mm&\omit&&\omit&&\omit&&\omit&&\omit&\cr
\noalign{\hrule} 
height1mm&\omit&&\omit&&\omit&&\omit&&\omit&\cr
&$\beta_P$&&\vphantom{\Big|}$2.55\pm0.01\;[2.53\pm0.01]$&&$2.50\pm0.08\;[2.55,\;\hbox{fix}]$&
&\hfil$2.56\pm0.01\;[2.56\pm0.01]$ \hfil&&
$2.56\pm0.03$& \cr 
\noalign{\hrule} 
height1mm&\omit&&\omit&&\omit&&\omit&&\omit&\cr
&\vphantom{\Big|}$\beta_{P'}$ 
\phantom{\big|}&&\phantom{\Big|}$1.04\pm0.02\;[1.09\pm0.02]$&&$1.46\pm0.17\;[1.04,\;\hbox{fix}]$&
&\hfil $1.04\pm0.02\;[1.04\pm0.02]$   \hfil&&\hfil 
$1.05\pm0.05$& \cr  
height1mm&\omit&&\omit&&\omit&&\omit&&\omit&\cr
\noalign{\hrule} 
height1mm&\omit&&\omit&&\omit&&\omit&&\omit&\cr
&\vphantom{\Big|}$\dfrac{\chi^2}{{\rm d.o.f.}}$ 
\phantom{\big|}&&\phantom{\Big|}$\dfrac{303}{229-4}\;\left[\dfrac{252}{229-4}\right]$&
&$\dfrac{109}{58-3}\;\left[\dfrac{45}{39-1}\right]$&
&\hfil $\dfrac{415}{288-5}\;\left[\dfrac{348}{268-5}\right]$  
\hfil&& & \cr  height1mm&\omit&&\omit&&\omit&&\omit&&\omit&\cr
\noalign{\hrule}}
\vskip.05cm}
\centerline{\box1}
\smallskip
\noindent{\petit$^{\rm (a)}$ We here endow all $\pi N$ numbers
 with a minimum systematic error of 1.5\%.}
\noindent{\petit$^{\rm (b)}$ By ``cut" we mean that $\pi\pi$ data 
for $s^{1/2}<2\,\gev$    
are removed from the fit. $^{\rm (c)}$ We here fix $\beta_P$, $\beta_{P'}$ 
as given by $NN$, $\pi N$ to avoid spureous minima.\hb
$^{(*)}$ The error in this quantity will be improved using crossing sum rules; 
see Eq.~(11) below.}
\medskip
\centerline{\sc Table~I}
\centerrule{5truecm}
\bigskip}
\centerline{\box0}
}\endinsert

 A few features of our results worth noting are the following. 
Firstly,  the equality of $f_{N/\pi}$ and $\beta_P$, $\beta_{P'}$,  for fits with and 
without $\pi\pi$ data is a very satisfactory test of factorization. 
Another interesting point is the stability and accuracy of the parameters 
$f_{N/\pi}$,  $\beta^{(N\pi)}_\rho$, $\beta_P$. 
The parameter 
 $\beta_{P'}$ is less well determined, and $\beta_\rho$
 is not fixed with precision by  fits to data alone; 
we will improve its accuracy in a moment using sum rules.
Secondly, the matching between the low energy ($s^{1/2}\leq1.42\,\gev$) 
results for
cross sections from phase shift analyses, and the high energy  
($s^{1/2}\geq1.42\,\gev$)  
Regge representations is excellent for $\pi^0\pi^-$, $\pi^-\pi^-$ and 
$\sigma^{(I_t=0)}$.
It is less good for $\pi^+\pi^-$, where matching occurs only at the 
$1.5\,\sigma$
level, no doubt due to the coinciding tails of
 the $f_2(1270)$ and $f_0(1370)$
resonances.  
And, thirdly, the fact that, for $NN$ and $\pi N$ the \chidof\ 
is somewhat larger than unity is due to the following effects. 
First, we use only two poles for vacuum exchange, and one for 
charge exchange: 
we are thus missing the contributions of other
poles, likely small, but not negligible at the lower energy range. 
Secondly,   at the very low
energy range,  the experimental cross sections oscillate a little around  the Regge formulas. 
Finally, and probably the most important effect, we have  that,
 to cover well the upper part of
the  energy range, we need more sophisticated  expressions: see \sect~4. 

Besides this, we have a few technical points to make in connection with the fits including 
$\pi\pi$ data. As is clear from \fig~3, the low energy ($s^{1/2}<2.5\,\gev$) 
results for
$\pi^-\pi^-$ cross sections of various experiments 
are quite incompatible with one another, which is the reason 
for the large \chidof\ in {\sl no-cut} fits. 
There is certainly  
a bias in the experimental  
 $\pi^-\pi^-$ 
 cross sections of Biswas et al., and Robertson, Walker and Davis,\ref{7} 
in the 
lower energy range. 
This is probably due to incorrect treatment of final state interactions, 
that, at these lower energies, are influenced by the $\Delta_{33}$ and other 
resonances.  
At higher energies the influence of this 
resonance seems to become negligible as, indeed, the 
 $\pi^-\pi^-$  cross sections found by 
Robertson, Walker and Davis overlap those of Abramowicz et al.\ref{7} and both 
  tend to the 
$\pi^+\pi^-$  one, as Regge theory  and the Pomeranchuk theorem imply. 
We consider that this problem is solved by considering our two types of fits, {\sl cut}
 or {\sl no-cut},
 for $\pi\pi$ scattering. 
 
\topinsert{
\medskip
\centerline{
\epsfxsize 10.truecm\epsfbox{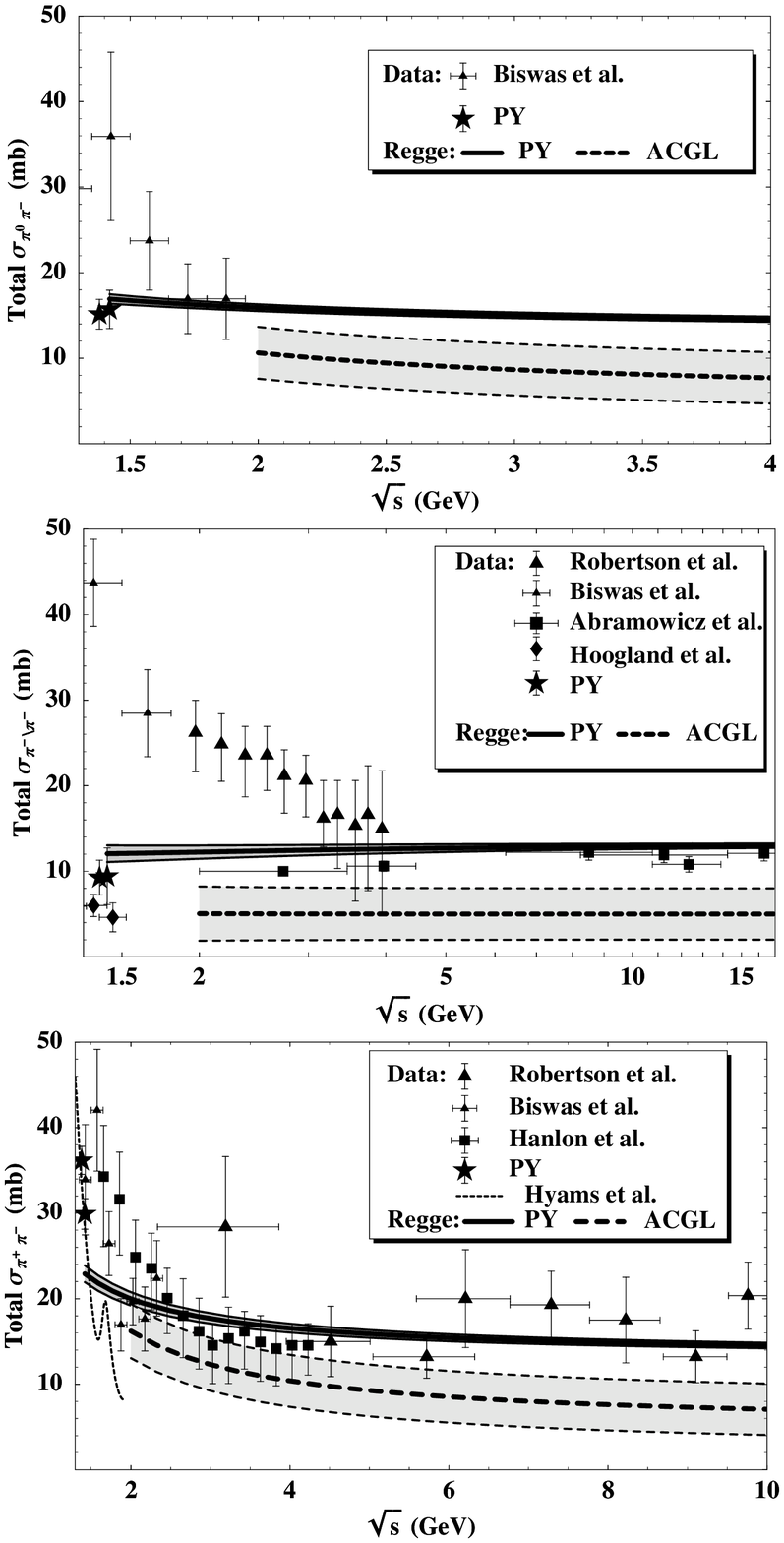}}
\bigskip
\setbox5=\vbox{\hsize 14.6truecm\captiontype\figurasc{Figure 3. }
 {Total cross sections
$\sigma_{\pi^0\pi^-}$, $\sigma_{\pi^-\pi^-}$ and $\sigma_{\pi^+\pi^-}$. 
Black dots, triangles and squares: experimental points from ref.~7. 
The stars at 1.38 and 1.42 \gev\ (PY) are from the phase shift analysis 
of experimental data given in ref.~11, slightly improved for the D2 wave. 
Continuous lines, from 1.42 \gev\ (PY): Regge formula, with parameters as in 
our best fit 
(the three lines per fit    cover the error in the theoretical values of the 
Regge residues). 
Dashed lines, above 2 \gev: the cross sections following from ACGL;\ref{1} 
the grey band covers their error band.
Below 2 \gev, the dotted line 
corresponds to  the $\pi^+\pi^-$ cross section from 
 the Cern--Munich analysis; cf.~\fig~7 in the paper of Hyams~et~al.\ref{8}}}
\centerline{\box5}
\bigskip
}\endinsert

We next discuss the  isospin~2 exchange piece, $R_2(s,t)$. 
We have three methods to get the quantity  $\beta_2$. 
First,  
we  fix the values of  $\beta_P$ and $\beta_{P'}$ 
to their best values, as given in Table~I, and fit the 
$\pi\pi$ data using Eqs.~(4), (5). 
Note that one cannot 
leave the parameters   $\beta_P$, $\beta_{P'}$ free 
in these fits because one would get spureous minima, 
since the data are not precise enough. 
We find $\beta_\rho=1.07$ and a very small $\beta_2\sim-2\times10^{-8}$. 
Alternatively, we could obtain  $\beta_2$ by fitting 
 $\sigma_{\pi^0\pi^0}-\sigma_{\pi^0\pi^+}$ at $s^{1/2}=1.42\,\gev$, as was done in ref.~11. 
This gives 
$\beta_2=0.55\pm0.2$. Finally, we can use the first crossing sum rule 
in the Appendix to ref.~11 (identical to (B.7) in ACGL), which would give a
$\beta_2$ compatible with zero. 
We take as a compromise the number
$$\beta_2=0.2\pm0.2.
\equn{(7)}$$
However, we should note that the $t$ dependence of  $R_2(s,t)$
is little more than guesswork.

\medskip
\noindent{\sl Sum rules.}\quad We now say a few words on 
 the sum rules discussed in ref.~11. 
Because these sum rules were verified with  Regge expressions slightly different from 
what we have now found, 
one may wonder what happens to them. 
Since the formulas in (4), (5),  
with parameters as in Table~I, agree with those of ref.~11 within $\lsim2\,\sigma$, 
and the decrease of $\beta_P$ is (partially) 
compensated by the increase in  $\beta_{P'}$,
 it can be expected that the 
various sum rules would still be satisfied within errors,
 as indeed it happens. 
Our numbers here leave the agreement of the Olsson sum rule
 and the value of the P wave
scattering length and effective range  still within $1\,\sigma$. 
We have already discussed the first crossing sum rule  in the Appendix in ref.~11 
in connection with $\beta_2$, so 
we turn to the second crossing sum rule.  It reads, 
$$ \int_{4m^2_\pi}^\infty\dd s\,\dfrac{\imag F^{(I_t=1)}(s,4m^2_\pi)-\imag F^{(I_t=1)}(s,0)}{s^2}=
 \int_{4m^2_\pi}^\infty\dd s\,\dfrac{8m^2_\pi[s-2m^2_\pi]}{s^2(s-4m^2_\pi)^2}
\imag F^{(I_s=1)}(s,0).
\equn{(8)}$$
The interest of this sum rule lies in that its high energy ($s^{1/2}\geq1.42\,\gev$) 
is dominated by $\rho(s,t)$, while the low energy piece ($s^{1/2}\leq1.42\,\gev$) 
is such that 
the contributions of the S waves cancel, so it is dominated by the  P wave, 
which is very well known. 
Thus, it provides an independent, reliable way of fixing the parameter $\beta_\rho$. 
We find (8) satisfied provided one has 
$$\beta_\rho=0.82\pm0.12.
\equn{(9)}$$
Since this is compatible with the independent determinations in Table~I, 
  we may include fulfillment of (8) in the fits. 
If we do so for the fit with {\sl cut} $\pi\pi$ data, we get the  value
$$\beta_\rho=0.78\pm0.11.
\equn{(10a)}$$
If we include (8) in the fit with all $\pi\pi$ data ({\sl no-cut}) 
we find instead
$$\beta_\rho=1.07\pm0.09.
\equn{(10b)}$$
Combining (10a,b) we can then take
$$\beta_\rho=0.94\pm0.10\,\hbox{(Stat.)}\pm0.10\,\hbox{(Syst.)}.
\equn{(10c)}$$

\smallskip
\noindent{\sl Best values}.\quad
We can now present our best values, and compare them 
 with the
values given in ref.~11 (PY), obtained  basically from
 those by Rarita et al.,\ref{13}  or those of refs.~1,~5 (ACGL):
$$\matrix{&\hbox{[Our best values]}\quad&\phantom{jupii!}&\hbox{[PY]}\quad&\hbox{[ACGL].}\cr
\beta_\rho&0.94\pm0.14&
\phantom{yupii!}&0.84\pm0.10&1.48\pm0.25\cr
\beta_P&2.56\pm0.03&\phantom{yupii!}&3.0\pm0.3&1.0\pm0.6\cr
\beta_{P'}&1.05\pm0.05&\phantom{yupii!}&0.72\pm0.07&2.22\pm0.38\cr
\beta_2&0.2\pm0.2&\phantom{yupii!}&0.55\pm0.20&0\cr
}
\equn{(11a)}$$
Besides these, we have also 
$$f_{N/\pi}=1.408\pm0.005,\quad\beta^{(N\pi)}_\rho=0.366\pm0.010.
\equn{(11b)}$$

Our present results are 
 compatible with those 
in refs.~6,~11,~13. 
We note, however, that our fits include much more information on the total cross sections
 than those 
in refs.~6,~13. The first only includes $\pi^+\pi^-$ data while the 
more complete
 fit of Rarita et al.\ref{13}  includes 24 total cross section data for 
$NN$ (we have 34) and 28 for $\pi N$ (we have 87); the energy range we cover is also wider, 
by a 
factor 6 in the variable $s$. 
We also have 58 $\pi\pi$ data points (none in ref.~13). 
Of course, the situation is different for the $t$ dependence of the residue functions $f_i(t)$ 
for which the fit of Rarita et al.\ref{13} cannot be really improved. 

The results in (11) may be compared with some theoretical models. The value
$f_{N/\pi}\simeq1.4$ is similar to what one gets in the 
 naive quark model\ref{17} with additive quark-quark cross sections, that gives $f_{N/\pi}=3/2$. 
[It is, however, 
not clear why the naive quark model  works, as its mechanism is very different 
from the orthodox QCD one]. 
Likewise, the value of $\beta_\rho=0.94\pm0.14$ is similar to what one has in the 
Veneziano model\ref{18} ($\beta_\rho\simeq0.95$), but 
the relation $\beta_{P'}=\tfrac{3}{2}\beta_\rho$ 
that this model gives is not well satisfied.
$\beta_\rho$ also agrees with the rho dominance model, in which one couples the rho universally to
pions and nucleons according to
$$g\,\bar{N}\vec{t}\gamma^\mu N \vec{\rho}_\mu,\quad
 g\,\left(\vec{\pi}\times\lrvec{\partial_{\mu}} \vec{\pi}\right)\vec{\rho}^{\,\mu}
$$
with $\vec{t}=\vec{\sigma}/2$, $\vec{\sigma}$ the Pauli matrices, 
that gives $\beta_\rho=\sqrt{\tfrac{8}{3}}\,f_{N/\pi}\,\beta^{(N\pi)}_\rho\simeq 0.84.$

\booksection{3. $\pi K$ scattering}

\noindent
The analysis of $\pi K$ scattering follows similar lines. For exchange of 
isospin zero we have
$$\eqalign{ 
\imag F^{(I_t=0)}_{\pi K}(s,t)&\,\simeqsub_{{s\to\infty}\atop{t\,{\rm fixed}}}
f_{K/\pi}\Big[P(s,t)+rP'(s,t)\Big];\cr
f_{K/\pi}=&\,f_K^{(P)}(0)/f_{\pi}^{(P)}(0).\cr
}
\equn{(12a)}$$
$P$, $P'$ are as above, and $r$ is related to the branching ratio for the $\bar{K}K$ decay of the
resonance\fnote{Since the $P'$ pole couples so weakly to kaons, one may 
wonder on the importance of other Regge poles for
 the subleading contribution to kaon scattering. For $KK$ scattering, 
 the Regge pole associated with the $f_2(1525)$ resonance gives a substantial 
contribution; but, for 
$KN$ or $\pi K$ scattering, this trajectory contributes very little since it 
is almost uncoupled to pions and nucleons and its intercept is small, 
$\alpha_{f_2(1525)}\simeq-0.3$. 
For $KN$ and $\pi K$, the amplitude for exchange of  zero isospin is almost pure 
Pomeron.} 
$f_2(1270)$, $r\sim{\rm BR}=(4.6\pm0.5)\times10^{-2}$. 
 For isospin 1 exchange,
$$\eqalign{
\imag F^{(I_t=1)}_{\pi K}(s,t)&\,\simeqsub_{{s\to\infty}\atop{t\,{\rm fixed}}}
g_{K/\pi}\rho(s,t),\cr
g_{K/\pi}=&\,f_K^{(\rho)}(0)/f^{(\rho)}_\pi(0);\cr
}
\equn{(12b)}$$
$\rho(s,t)$ is as before. 
To find the desired representations for the $\pi K$ amplitude we have to determine 
the ratios $f_{K/\pi}$, $g_{K/\pi}$. 
For the first, this is done taking the $f_{N/\pi}$ from $NN$, $\pi N$
 scattering, as in the previous
sections,   and with the help of the 
 even combination of cross sections for  
$KN$ scattering:
$$\eqalign{
\sigma_{K^+ p}+\sigma_{K^- p}
\simeqsub_{s\;{\rm large}}&\;\dfrac{4\pi^2}{\lambda^{1/2}(s,m^2_K,m_p^2)}f_{N/\pi}f_{K/\pi}
\Big[P(s,0)+rP'(s,0)\Big].\cr}
\equn{(13)}$$
For $g_{K/\pi}$, unfortunately, we cannot use the charge exchange reaction 
$K^-p\to K^0n$ because there are two trajectories 
of comparable importance,  
$\rho$ and that corresponding to $a_2(1320)$ 
exchange, that contribute; for a discussion, cf.~for
 instance the text of Barger and Cline, ref.~12. 
The difference of cross sections $K^+p$ and $K^-p$ also contains 
extra contributions ($\omega$, $\phi$, \tdots).

For the $KN$ cross sections we will take data in the region 
$E_{\rm kin}>1\,\gev$, and go up to $E_{\rm kin}=10\,\gev$. 
At higher energies
 the logarithmic increase of the total 
cross section  for $K^+p$ scattering is noticeable,
 and we would need  more complicated Regge formulas (that we will give 
in  \sect~4); 
while, as occurs for the $\pi\pi$ case, the importance of 
 the very high energy region is negligible in most applications
 to $\pi K$ scattering. 
For $\pi K$ scattering we thus expect the ensuing Regge 
expressions to be accurately valid for a corresponding energy range, 
say, for $1.7\,\gev<s^{1/2}<11\,\gev$. 

The $K^\pm p$   data we take also from the COMPAS Group  compilations; see
 the Particle Data Tables.\ref{16} 
For those 
data where systematic errors are not given, we have included a common systematic error of 
 0.3~mb, as we did for the $\pi N$ case. 
We take only data at energies at which there
are  results for both $K^+p$ and $K^-p$. 
In the fits we use the very precise values of the 
parameters $f_{N/\pi}$, $\beta_P$ obtained before, 
and we set $r=0$, since it is very small and not very well known. 
In fact, in \sect~4 we will make fits leaving $r$ 
free; its value will turn out to satisfy $|r|\lsim0.1$.
We find
$$\eqalign{
f_{K/\pi}=&\,0.67\pm0.01\quad[{\rm from}\;K^+p+K^-p;\;\textstyle{\chidof}=50/(43-1)]\cr
g_{K/\pi}=&\,1.1\pm0.1.\cr
}
\equn{(14)}$$
The results for $(\sigma_{K^+ p}+\sigma_{K^- p})/2$ are shown in Fig.2. 
The  value of $g_{K/\pi}$ is taken from the classical analysis of ref.~19, that 
takes into account the $a_2(1320)$ exchange. 
The value of $f_{K/\pi}$ is within $20\%$ of its SU(3)
 value, $\sqrt{\tfrac{2}{3}}\simeq0.82$.

\booksection{4. A global fit valid up to multi TeV energies}  

\noindent
A simple parametrization of scattering amplitudes which fits data at energies 
$s^{1/2}>12$~GeV (with a $\chi^2/{\rm d.o.f.}=1.2$ to $1.8$, 
depending on the process) may be found in 
 in refs.~20,~21. Here the Pomeron is allowed an intercept larger than unity, 
$\alpha_P(0)\sim1.095$, and the intercept of the $P'$ is given as 
$\alpha_{P'}(0)=0.66$.
This parametrization, that we will call ``power Pomeron" parametrization,
 is purely phenomenological, as explicitly mentioned 
 in refs.~20,~21. 
Only data with energy larger than $\sim10\,\gev$ are used in the fits 
which,  
if extended to energies  below 5~GeV, miss  widely the data. 
These parametrizations also must fail at very large energies since 
they are incompatible with
unitarity in that  they violate the Froissart bound. 
As a matter of fact, in ref.~22 it is remarked the inadequacy of such parametrization, 
and a parametrization verifying the Froissart bound (i.e., with a term 
in ${\rm(Const.)}\times\log^2s/s_0+{\rm Const.}$) is substituted 
in place of the ``power Pomeron." 
This improves substantially the \chidof\ of the fit, and gives an intercept
$\alpha_{P'}(0)=0.54\pm0.02$, perfectly compatible with our choice $0.52\pm0.02$.
The corresponding parametrization holds down to $s^{1/2}=5\,\gev$.

It is  possible to write a parametrization, 
similar to that of ref.~22,  obtained by a modification of the Pomeron 
in Eq.~(4a), that fits data for kinetic energies from 1~GeV to 
the multi-TeV region and which, moreover, is compatible 
with unitarity by adding a slightly more complicated logarithmic term.  
We do this as follows: we note 
 that it is possible to improve the Froissart bound to a bound of the form\ref{23}
$$\sigma_{\rm tot}\leq a\log^2{{s}\over{s_1\log^{7/2}s/s_2}},
\eqno{(15)}$$
which is  maximal in the sense that one cannot increase the power of the 
log in the denominator to more than $\ffrac{7}{2}$. 
For the bound for $\pi\pi$ scattering, one can evaluate the constants $a$, $s_1,\,s_2$ 
in terms of the pion mass and
 low energy parameters for the D wave, with
$a=\pi/4m^2_\pi\simeq15\,{\rm mb}^2$, $s_1=m^2_\pi$
  if we assume the cross section to be mostly
inelastic.  What this suggests is that we add a term like (15) to the Pomeron given in (4a), 
but leaving  $a$, $s_1,\,s_2$ as free parameters. 
Thus we replace,
$$\eqalign{
P(s,t)=&\,\beta_P\,\alpha_P(t)\,{{1+\alpha_P(t)}\over{2}}\,
{\rm e}^{bt}(s/\hat{s})^{\alpha_P(t)}\to P_F(s,t),\cr
P_F(s,t)=&\,\left\{\widetilde{\beta}_P+A\log^2{{s}\over{s_1\log^{7/2}s/s_2}}\right\}\,
\alpha_P(t)\,{{1+\alpha_P(t)}\over{2}}\,
{\rm e}^{bt}(s/\hat{s})^{\alpha_P(t)}.\cr
}
\equn{(16)}$$
This replacement should also be made in Eqs.~(6), (12) and (13).
The logarithmic term has an appealing physical interpretation 
as the contribution of the Regge cuts which, as Mandelstam showed long ago,\ref{24} 
should accompany the Pomeron. 
The parameter $\beta_P$ that we used before is to be viewed as 
an effective parameter, the sum of $\widetilde{\beta}_P$ 
and the average value, for low energy ($s^{1/2}\lsim15\,\gev$), of 
the logarithmic piece in (16).

With (16) we  fit data for $\pi^\pm p$, $K^+p+K^-p$, $\pi\pi$ 
 and $pp+\bar{p}p$   cross
sections\fnote{Above  30~GeV we approximate 
$\sigma_{\bar{p}p}-\sigma_{pp}=(66.7\,{\rm mb})(s/\hat{s})^{-0.55}$,  
where this difference comes
from the phenomenological fit of ref.~16, since  we do not have data at coinciding energies. 
For $\pi\pi$ only data above 2 \gev\ are included in these fits.}
  up
to the highest energies attained experimentally, 
30~TeV in cosmic ray experiments.\ref{25}

Because we have so many experimental data, covering such a
wide energy range, we may fit all hadronic data (i.e.,
including $NN$, $\pi N$, 
$KN$ and $\pi\pi$ data)  leaving all parameters free; 
in particular, this will test the quality of the assumption of degenerate 
rho and $f_2$ trajectories, the equality of  $f^{(P)}_{N/\pi}$,  $f^{(P')}_{N/\pi}$,
 and the
smallness of the parameter $r$  in Eq.~(13). 
We only fix $\alpha_\rho(0)$ to the number given by 
deep inelastic scattering, $0.52\pm0.02$,  include the sum rule (8), and find
$$\eqalign{
f^{(P)}_{N/\pi}=&\,1.350\pm0.008;\quad f^{(P')}_{N/\pi}=1.67\pm0.07;\quad 
f_{K/\pi}=0.74\pm0.01;
\quad
\alpha_{P'}(0)=0.61\pm0.04,\cr
\widetilde{\beta}_P=&\,2.33\pm0.09;\quad
\beta_{P'}=1.05\pm0.10;
\quad\beta^{(N\pi)}_\rho=0.385\pm0.009;\quad \beta_\rho=0.94\pm0.14;\quad r=0.11\pm0.08,\cr
A=&\,0.022\pm0.002;\quad
 s_1=(1.2\pm0.7)\times 10^{-4}\;{\gev}^2;\quad
s_2=(0.33^{+0.6}_{-0.3})\times10^{-7}\;{\gev}^2,\cr
\chi^2&\,/({\rm d.o.f.})=339/(358-12)\simeq0.98. 
\cr}
\equn{(17)}$$

\topinsert{
\setbox0=\vbox{\hsize16truecm{\epsfxsize 13.0truecm\epsfbox{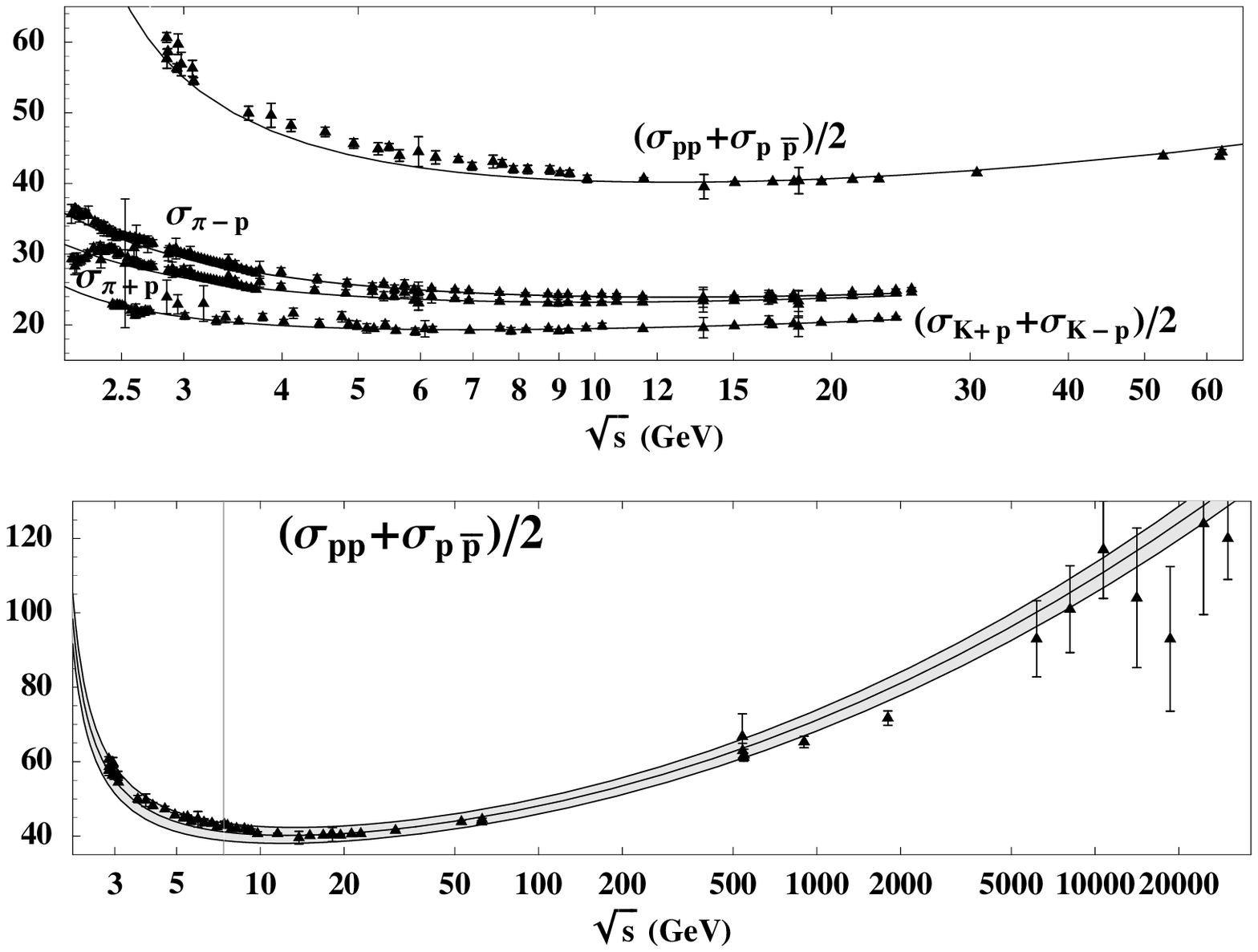}}}
\centerline{{\box0}}
\bigskip
\setbox1=\vbox{\hsize 14.5truecm\captiontype\figurasc{Figure 4. }
 {The total cross sections $\sigma_{\pi^\pm p}$, 
$\tfrac{1}{2}(\sigma_{K^+p}+\sigma_{K^-p})$ and  
 $\tfrac{1}{2}(\sigma_{\bar{p}p}+\sigma_{pp})$ up to $30\,-\,60\;\gev$ 
(upper graph) and  $\tfrac{1}{2}(\sigma_{\bar{p}p}+\sigma_{pp})$ 
up to 30~TeV (lower graph). 
Black dots, triangles and squares: experimental points. 
For energies above 30 \gev, we have given the experimental values of 
 $\tfrac{1}{2}(\sigma_{\bar{p}p}+\sigma_{pp})$ as if they equaled $\sigma_{\bar{p}p}$
 or $\sigma_{pp}$.
Continuous lines: Regge formulas, with parameters as in  (18). 
In the lower figure we have given the error bands 
for  $\tfrac{1}{2}(\sigma_{\bar{p}p}+\sigma_{pp})$  that follow from (18).}}
\centerline{\box1}
}\endinsert

The value of $\beta_\rho$ given here is that found before, Eq.~(10c); 
since there are no  $\pi\pi$ data at very high energy, the value of this 
quantity essentially decouples from the very high energy analysis.\fnote{If we had fitted 
also $\beta_\rho$, including the sum rule (8), its value would 
depend on whether we had included all $\pi\pi$ data above 1.4 \gev\
 (in which case we would have got $1.05\pm0.009$) 
or only data for $s^{1/2}\geq2\,\gev$, which gives $0.80\pm0.11$: 
 essentially the same numbers as in the fits in \sect~2, 
Eq.~(10a,b).}
 
What is interesting of (17) is that the value of $\alpha_{P'}(0)$ is compatible 
with what one finds from degeneracy, $\alpha_{P'}(0)=\alpha_\rho(0)=0.52\pm0.02,$ 
and that $f^{(P)}_{N/\pi}$ and $f^{(P')}_{N/\pi}$ are not far from each other, as required by 
(strong) factorization. In fact, this had alredy been noticed in 
ref.~22: in a fit with a formula compatible with theory (the Froissart 
bound), the resuts respect other theoretical constraints automatically.  
The problem with the fit in (17) is that   
there is, unfortunately, a very strong correlation among $\widetilde{\beta}_P$, $\beta_{P'}$, 
$\alpha_{P'}(0)$, $s_1$ 
and $s_2$ and, if we leave all of them  free as we did in getting (17),
 there exist   
a large number of equally significant minima: 
 the parameters are not well determined. 
In fact, 
$s_1$, $s_2$, $\beta_{P'}$ and $\alpha_{P'}(0)$ can one mock the effects of each other.
In particular, a set of fits with quality essentially unchanged may be 
obtained by varying simultaneously $s_1$ and $s_2$.
In view of this, we require $f^{(P)}_{N/\pi}=f^{(P')}_{N/\pi}$ and
, to fix the parameters, choose
 $s_1=0.01\,\gev^2$ and repeat the fit with 
all other parameters free. 
We  find what we consider our best result: 
$$\eqalign{
f_{N/\pi}=&\,1.380\pm0.004\quad f_{K/\pi}=0.717\pm0.005;\quad
\alpha_{P'}(0)=0.55\pm0.03;
\quad r=0\pm0.013,\cr
\widetilde{\beta}_P=&\,2.31\pm0.05;\quad
\beta_{P'}=1.39\pm0.14;
\quad\beta^{(N\pi)}_\rho=0.377\pm0.009;\quad \beta_\rho=0.94\pm0.14,\cr
A=&\,0.033\pm0.001;\quad
 s_1=0.001\;{\gev}^2[{\rm fix.}];\quad
s_2=0.13\pm0.05\;{\gev}^2;\cr
\chi^2&\,/({\rm d.o.f.})=372/(358-10)\simeq1.066. 
\cr}
\equn{(18)}$$
We note that, although the \chidof\ is slightly 
worse than that in (17), we consider the fit in 
(18) to be more satisfactory physically. The values of 
the parameters $s_1$, $s_2$ in (17) were too
small 
for comfort, and  one should not force too good a fit at the expense of 
physical considerations (like factorization or degeneracy), particularly  
since we are fitting with formulas that, at the lowest energies, should be corrected 
by including other Regge poles (or cuts). 
Eq.~(18) has the nice properties that degeneracy is verified, up to errors, 
that $f_{K/\pi}$ agrees better with its SU(3) value, and the 
value of $\beta_{P'}$ agrees, also within errors, with the prediction of the Veneziano model, 
$\beta_{P'}=\tfrac{3}{2}\beta_\rho=1.4\pm0.3$.

At the lower energies (below 15~GeV)
 (16) plus (17) or (18) overlap with the previous fits, using (4a) for the Pomeron 
and $P'$, for vacuum exchange.
In fact, for
$Kp$ or
$\pi N$, the corresponding curves could not  be distinguished from those obtained using (4a) 
in Fig.~2; see \fig~4. 
For $\bar{p}p+pp$, the result of the fits with 
the two types of formulas, (4a) and  (18) are depicted in Fig.~4, 
where the error bars corresponding to (18) are also  shown.

\booksection{5. Summary, and a short discussion}

\noindent
The Regge parameters that 
ACGL\ref{1} and, following them, the authors in refs.~2,~3,~4,~17
 assume  not only are unorthodox but, as we have shown, incompatible 
with experiment. As our \fig~3 clearly demonstrates, the claimed large
 errors in ACGL are not large enough to
cover the experimental data. 

ACGL get these quaint Regge parameters by considering sum rules like 
(8) that link the Regge contributions, 
which they assume to hold only 
 for $s^{1/2}\geq 2\gev$,  
 with the corresponding low energy ($s^{1/2}< 2\gev$) pieces. 
Unfortunately, the intermediate
intermediate energy ($1.4\,\gev\leq s^{1/2}<2\,\gev$) that 
 ACGL, again here followed by the authors in
refs.~2,~3,~4, 
 take for the S0, P, D0 and F phases  come basically
 from 
 the experimental analysis of
the Cern--Munich group, 
whose $\pi^+\pi^-$ cross section is  more and more incompatible, as $s^{1/2}$ nears 2 \gev\
--in fact, as soon as inelasticity becomes
important-- with 
the values found by all other experiments:\ref{7} see our Fig.~3. 
[The interested reader may consult 
ref.~9 for the detailed discussion of this and other related issues].
  It is thus not surprising that
Pennington\ref{5} and Ananthanarayan et al.,\ref{1} who 
fix their Regge parameters by balancing them above $2\,\gev$ with 
phase shifts below $2\,\gev$, get totally incorrect Regge amplitudes. 
And, given these facts, 
it also follows that  the {\sl low energy} results of  references~2,~3,~4, 
which borrow their input at energies  $s^{1/2}\geq1.4\,\gev$ from
ACGL,   should be taken with great caution.

\midinsert{
\setbox1=\vbox{\petit \offinterlineskip\hrule
\halign{
&\vrule#&\strut\hfil\quad#\quad\hfil&\vrule#&\strut\hfil\quad#\quad\hfil&
\vrule#&\strut\hfil\quad#\quad\hfil&\vrule#&\strut\hfil\quad#\quad\hfil\cr
 height2mm&\omit&&\omit&&\omit&&\omit&\cr 
&\hfil \hfil&&\hfil $1\,\gev\lsim  E_{\rm kin}\lsim15\,\gev$\hfil&
&\hfil ${{\hbox{$1\,\gev\lsim E_{\rm kin}\lsim30$~TeV}}\atop{\hbox{all parameters free}}}$ \hfil&
&\hfil ${{\hbox{$1\,\gev\lsim  E_{\rm kin}\lsim30$~TeV}}\atop\hbox{$s_1=0.01,\,
f^{(P)}_{N/\pi}=f^{(P')}_{N/\pi}$.}}$\hfil&
\cr
 height1mm&\omit&&\omit&&\omit&&\omit&\cr
\noalign{\hrule} 
height1mm&\omit&&\omit&&\omit&&\omit&\cr
&$f^{(P)}_{N/\pi}$&&\vphantom{\Big|}$1.408\pm0.005$&
&\hfil $1.350\pm0.008$\hfil&&
$1.380\pm0.004$& \cr
\noalign{\hrule} 
height1mm&\omit&&\omit&&\omit&&\omit&\cr
&\vphantom{\Big|}$f^{(P')}_{N/\pi}$ 
\phantom{\big|}&&\phantom{\Big|}$\equiv f^{(P)}_{N/\pi}$ [fix]&
&\hfil $1.67\pm0.07$ \hfil&&$\equiv f^{(P)}_{N/\pi}$ [fix] & \cr
\noalign{\hrule} 
height1mm&\omit&&\omit&&\omit&&\omit&\cr
&$f_{K/\pi}$&&\vphantom{\Big|}$0.67\pm0.01$&
&\hfil $0.74\pm0.01$\hfil&&$0.717\pm0.005$& \cr
\noalign{\hrule} 
height1mm&\omit&&\omit&&\omit&&\omit&\cr
&\vphantom{\Big|}$r$&&$0$ [fix]&
&\hfil $0.11\pm0.08$\hfil&&$0\pm0.013$& \cr
\noalign{\hrule}
height1mm&\omit&&\omit&&\omit&&\omit&\cr
&\vphantom{\Big|}$\alpha_\rho(0)$&
&$0.52\pm0.02$ [fix]&&\hfil $0.52\pm0.02$ [fix]\hfil&&$0.52\pm0.02$ [fix]& \cr
\noalign{\hrule}
height1mm&\omit&&\omit&&\omit&&\omit&\cr
&\vphantom{\Big|}$\alpha_{P'}(0)$&
&$0.52\pm0.02$ [fix]&&\hfil $0.61\pm0.04$\hfil&&$0.55\pm0.03$& \cr
\noalign{\hrule}
height1mm&\omit&&\omit&&\omit&&\omit&\cr
&\vphantom{\Big|}$\widetilde{\beta}_P$&&--&
&\hfil  $2.33\pm0.09$\hfil&&$2.31\pm0.05$& \cr
 height1mm&\omit&&\omit&&\omit&&\omit&\cr
\noalign{\hrule}
height1mm&\omit&&\omit&&\omit&&\omit&\cr
&\vphantom{\Big|}${\beta}_P$ &&\hfil$2.56\pm0.03$\hfil &&\hfil -- \hfil&&
--&
\cr
\noalign{\hrule}height1mm&\omit&&\omit&&\omit&&\omit&\cr
&\vphantom{\Big|}${\beta}_{P'}$&&$1.05\pm0.05$&
&\hfil $1.05\pm0.10$ \hfil&&$1.39\pm0.14$& \cr
\noalign{\hrule}height1mm&\omit&&\omit&&\omit&&\omit&\cr
&\vphantom{\Big|}$\beta^{(N\pi)}_\rho$ &&$0.366\pm0.010$&
&\hfil $0.385\pm0.009$ \hfil&&$0.377\pm0.009$& \cr
\noalign{\hrule}height1mm&\omit&&\omit&&\omit&&\omit&\cr
&\vphantom{\Big|}$\beta_\rho$ &&$0.94\pm0.14$&
&\hfil  $0.94\pm0.14$ [fix] \hfil&&$0.94\pm0.14$ [fix]& \cr
\noalign{\hrule}height1mm&\omit&&\omit&&\omit&&\omit&\cr
&\vphantom{\Big|}$A$&&\hfil --\hfil&
&\hfil $0.022\pm0.002$ \hfil&&$0.033\pm0.001$& \cr
\noalign{\hrule}height1mm&\omit&&\omit&&\omit&&\omit&\cr
&\vphantom{\Big|}$s_1$&&\hfil --\hfil&
&\hfil $(1.2\pm0.7)\times10^{-4}\,\gev^2$ \hfil&&$\equiv0.01\,\gev^2$& \cr
\noalign{\hrule}height1mm&\omit&&\omit&&\omit&&\omit&\cr
&\vphantom{\Big|}$s_2$&&\hfil --\hfil&
&\hfil  $(0.33^{+0.6}_{-0.3})\times10^{-7}\,\gev^2$ \hfil&&$0.13\pm0.05\,\gev^2$& \cr
\noalign{\hrule}height1mm&\omit&&\omit&&\omit&&\omit&\cr
&\vphantom{\Big|}\chidof&&\hfil\hfil --\hfil\hfil&&\hfil $0.98$ \hfil&&$1.07$& \cr
\noalign{\hrule}}
\vskip.05cm}
\centerline{\box1}
\smallskip
\centerline{\sc Table~II}
\centerrule{6cm}
\medskip
}\endinsert
 
Unlike the results of phase shift analyses, 
 the Regge formulas in Eqs.~(4), (16)
with the parameters as the ``Best values" in (11) or (18), and which we summarize in Table~II,
 give
a  consistent representation for  
the imaginary part of all the $\pi\pi$ scattering amplitudes, 
a representation which 
can be trusted, within the given errors, for $s^{1/2}>1.4$ \gev, provided  
$|t|^{1/2}<0.4$~\gev. 
In fact, one has better than that: 
our Regge formulas give a good representation 
 of those processes in pion-pion scattering where resonances are absent, or 
are not important, down to lower energies, just as it happens in $NN$ or $\pi N$
scattering. This occurs, in particular,
 for $\pi^0\pi^+$ and $\pi^-\pi^-$, for which the Regge formulas
reproduce the experimental data down to
$s^{1/2}\sim1.1\,\gev$. 
However, by the very nature of things, we are likely to   
have  uncertainties of the order of 15\% in the region 
$1.4\,\gev\leq s^{1/2}\leq1.8\,\gev$  when exchange of
 isospin~1 is important, because the Regge formula  probably represents data only 
in the average there, as occurs for $\pi N$ scattering.
Finally, and using Eqs.~6,~13 and the formulas in the last column in Table~II, we can fit 
$NN$, $\pi N$ and $KN$
 up to multi TeV energies, and predict  $\pi\pi$ and $\pi K$ cross sections there.

When performing calculations of $\pi\pi$ scattering in which 
the lower energy region is dominant (such as Roy equations, dispersion relations or sum rules) 
it is irrelevant, within our errors, which form one uses for the Pomeron, (4a), (17) or (18). 
The last has better overall fit, and (probably) a 
more realistic value for $\beta_{P'}$;
although the first is to be preferred in that it is simpler and fits slightly better 
the low energy data. 
The safest procedure is to use both fits, and consider their difference as a measure of the
influence of the parametrization in the results. 
We should, however, emphasize that the parameters in the fits are strongly correlated and, 
even when they are similar, one {\sl cannot} 
mix parameters from the various columns in Table~II; 
each fit stands on its own.

One may also wonder what happens for values of the momentum transfer larger than 
$|t|^{1/2}\sim0.4\,\gev$. On general grounds, one expects Regge theory to work 
when $s\gg\lambdav^2$, $s\gg |t|$ and in fact, as already mentioned, Regge 
representations for $NN$ or $\pi N$ become unreliable at large $|t|$. 
For example, the parametrizations of Rarita et al.\ref{13} and ref.~6 
for $f^{(\rho)}(t)$
differ completely from one another  already at $-t=0.23\,\gev^2$, where the 
first changes sign.  
There is unfortunately no sure way out of this problem 
(which is further discussed in the second paper in ref.~23), and 
one has to admit that, for $s^{1/2}>1.4\,\gev$ and  values of the momentum transfer 
$|t|>0.15\,\gev^2$, 
there is no reliable information on the pion-pion scattering amplitude 
--which, in particular, is an unavoidable cause of  uncertainty for 
 Roy equation analyses that require information for values of $|t|$ as large as 
$0.5\,\gev^2$.


\brochuresection{Acknowledgments}
\noindent
We are grateful to CICYT, Spain, and to INTAS, for partial financial support.
J.R. Pel\'aez thanks partial support from  
the Spanish CICYT projects,
BFM2000-1326 and BFM2002-01003 and the 
E.U. EURIDICE network contract no. HPRN-CT-2002-00311.

\brochuresection{References}
\item{1 }{Ananthanarayan, B., et al., {\sl Phys. Rep.}, {\bf 353}, 207,  (2001).}
\item{2 }{Colangelo, G., Gasser, J.,  and Leutwyler, H.,
 {\sl Nucl. Phys.} {\bf B603},  125, (2001).}
\item{3 }{Descotes, S., Fuchs, N. H.,  Girlanda, L., and   Stern, J., {Eur. Phys. J. C}, 
{\bf 24}, 469, (2002); Kami\'nski, R., Le\'sniak, L., and Loiseau, B. {\sl Phys. Letters} {\bf B551},
241 (2003).}
\item{4 }{Buttiker, P., Descotes-Genan,~S., and Moussallam,~B., LPT-Orsay-03-76 
(hep-ph/0310283).}
\item{5 }{Pennington, M. R., {\sl Ann. Phys.} (N.Y.), {\bf 92}, 164, (1990).}
\item{6 }{Froggatt,~C.~D., and Petersen,~J.~L., {\sl Nucl. Phys.} {\bf B129}, 89 (1977).}
\item{7 }{Biswas, N. N., et al., {\sl Phys. Rev. Letters}, 
{\bf 18}, 273 (1967) [$\pi^-\pi^-$, $\pi^+\pi^-$ and $\pi^0\pi^-$];
 Cohen, D. et al., {\sl Phys. Rev.}
{\bf D7}, 661  (1973) [$\pi^-\pi^-$];
 Robertson, W. J.,
Walker, W. D., and Davis, J. L., {\sl Phys. Rev.} {\bf D7}, 2554  (1973)  [$\pi^+\pi^-$]; 
Hoogland, W., et al.  {\sl Nucl. Phys.}, {\bf B126}, 109 (1977) [$\pi^-\pi^-$];
Hanlon, J., et al,  {\sl Phys. Rev. Letters}, 
{\bf 37}, 967 (1976) [$\pi^+\pi^-$]; Abramowicz, H., et al. {\sl Nucl. Phys.}, 
{\bf B166}, 62 (1980) [$\pi^+\pi^-$]. These  references cover the region
between  1.35 and 16 \gev, and agree within errors in the regions where they overlap 
(with the exception of $\pi^-\pi^-$ below 2.3 \gev, see text).}
\item{8 }{Hyams, B., et al., {\sl Nucl. Phys.} {\bf B64}, 134, (1973); 
Grayer, G., et al.,  {\sl Nucl. Phys.}  {\bf B75}, 189, (1974). See also the analysis of the 
same experimental data in
Estabrooks, P., and Martin, A. D., {\sl Nucl. Physics}, {\bf B79}, 301,  (1974) 
and Au,~K.~L., Morgan,~D., and Pennington,~M.~R. {\sl Phys. Rev.} {\bf D35}, 1633 (1987).}
\item{9 }{Yndur\'ain, F. J., FTUAM 03-14
(hep-ph/0310206).}
\item{10 }{Yndur\'ain, F. J., {\sl Phys. Letters} {\bf B578},
99 (2004).}
\item{11 }{Pel\'aez, J. R., and Yndur\'ain, F. J., {\sl Phys. Rev.} {\bf D68}, 074005 (2003).}
\item{12 }{Gell-Mann, M.  {\sl Phys. Rev. Letters}, {\bf 8}, 263, (1962); 
Gribov, V. N., and Pomeranchuk, I. Ya.  {\sl Phys. Rev. Letters}, {\bf 8}, 343,  (1962). 
For more references in general 
Regge theory, see 
 Barger, V. D., and
Cline,~D.~B., {\sl Phenomenological Theories of High Energy  Scattering}, Benjamin, New~York, 
(1969). For references to the QCD  analysis, cf. 
Gribov, V. N., and  Lipatov, L. N. {\sl Sov. J. Nucl. Phys.} {\bf 15}, 438 and 675 
(1972); 
Kuraev, E. A., Lipatov, L. N., and Fadin, V. S.
 {\sl Sov. Phys. JETP} {\bf 44}, 443 (1976); 
Dokshitzer,~Yu.~L.  {\sl Sov. Phys. JETP} {\bf 46,} 641 (1977); 
Balitskii, Ya. Ya., and  Lipatov, L. N.  {\sl  Sov. J. Nucl. Phys.} {\bf 28}, 822 (1978);
Altarelli, G., and Parisi, G. {\sl Nucl. Phys.} {\bf B126}, 298  (1977). 
For phenomenological Regge analysis of deep inelastic scattering, see Martin, F. {\sl Phys.
Rev.} {\bf D19}, 1382 (1979); L\'opez,~C., and Yndur\'ain,~F.~J.,
 {\sl Nucl. Phys.}   {\bf B171}, 231 (1980); 
Adel,~K., Barreiro,~F., and Yndur\'ain, F.~J., {\sl Nucl. Phys.} 
{\bf B495}, 221 (1997). 
For the QCD theory of Regge trajectories, see Dubin,~A.~Yu., Kaidalov,~A.~B., and 
Simonov,~Yu.~A., {\sl Phys. Letters}, {\bf B323}, 41 (1994); Simonov,~Yu.~A., {\sl Nucl.~Phys.},
{\bf B324}, 67 (1989), 
from which earlier references may be traced.}
\item{13 }{Rarita, W., et al., {\sl Phys. Rev.} {\bf 165}, 1615, (1968).}
\item{14 }{Brodsky, S. J., and Farrar, G., {\sl Phys. Rev. Lett.}, {\bf 31}, 1153  (1973).}
\item{15 }{Barger, V., and Olsson,~M. {\sl Phys. Rev.}, {\bf 151}, 1123 (1966).}
\item{16 }{Particle Data Tables: Hagiwara, K., et al., {\sl Phys. Rev.} 
{\bf D66}, 010001 (2002).}
\item{17 }{Levin, H. J., and Frankfurter, L. L. (1965). {\sl JETP Letters} {\bf 2}, 65. 
For a comprehensive review, see Kokkedee, J. J. J. (1969),
 {\sl The Quark Model}, Benjamin, New
York.}
\item{18 }{Veneziano, G.  {\sl Nuovo Cimento} {\bf 57}, 190 (1968); Lovelace, C. {\sl Phys.
Letters} {\bf B28}, 264 (1968).}
\item{19 }{Barger, V., et al. {\sl Nucl. Phys.} {\bf B5}, 411 (1968).} 
\item{20 }{Cudell, J.~R., et al.,  {\sl Phys. Rev.} {\bf D61}, 034019 (2000).}
\item{21 }{Desgrolard, P., et al., {\sl Eur. Phys. J.} {\sl C18}, 555 (2001).}
\item{22 }{Cudell, J.~R., et al.,  {\sl Phys. Rev.} {\bf D65}, 074024 (2002).} 
\item{23 }{Yndur\'ain, F. J., {\sl Phys. Letters} {\bf B41},
591 (1972) and FTUAM 04-??, in preparation.}
\item{24 }{Mandelstam, S. {\sl Nuovo Cimento}, {\bf 30}, 1127, 1148 (1963).}
\item{25 }{Baltrusaitis, R. M.,  et al.  {\sl Phys. Rev. Lett.} {\bf 52}, 1380 (1984); 
Honda, M., et al. {\sl Phys. Rev. Lett.} {\bf 70}, 525 (1993).}

\bye